\documentclass[10pt,aps,pra,twocolumn,superscriptaddress,floatfix,longbibliography]{revtex4-2}
\usepackage[english]{babel}
\usepackage[utf8]{inputenc}
\usepackage{mathtools}
\usepackage{physics}
\usepackage{float}
\usepackage{braket}
\usepackage{bbm}
\usepackage{bm}
\usepackage{amsbsy}
\usepackage{amsthm}
\usepackage{amssymb}
\usepackage{amsfonts}
\usepackage{amsmath}
\usepackage{soul}
\setstcolor{red}
\usepackage{dsfont} 
\usepackage{graphicx} 
\usepackage{hhline}
\usepackage{epsfig}
\usepackage{epstopdf}
\usepackage{dsfont}
\usepackage{xcolor}
\usepackage{color}
\usepackage{verbatim}
\usepackage{nomencl}
\usepackage[colorlinks]{hyperref}
\usepackage{float}
\usepackage[normalem]{ulem}
\usepackage{multirow}
\usepackage{makecell}
\usepackage{url}
\usepackage{tikz}
\usetikzlibrary{arrows.meta, decorations.pathreplacing}
\DeclareUnicodeCharacter{2215}{/}
\DeclareUnicodeCharacter{0301}{\'{e}}

\begin{document} 
\title{Dynamics of Majorana zero modes across hybrid Kitaev chain}

\author{Rajiv Kumar}
\email{rajivkumar.rs.phy22@iitbhu.ac.in}
\affiliation{Department of Physics, Indian Institute of Technology (Banaras Hindu University) Varanasi - 221005, India}
\author{Rohit Kumar Shukla}
\email{rohitkrshukla.rs.phy17@iitbhu.ac.in}
\affiliation{Department of Chemistry; Institute of Nanotechnology and Advanced Materials; Center for Quantum Entanglement Science and Technology, Bar-Ilan University, Ramat-Gan, 5290002 Israel}
\author{Levan Chotorlishvili}
\email{levan.chotorlishvili@gmail.com}
\affiliation{Department of Physics and Medical Engineering, Rzeszow University of Technology, 35-959 Rzeszow Poland}
\author{Sunil K. Mishra}
\email{sunilkm.app@iitbhu.ac.in}
\affiliation{Department of Physics, Indian Institute of Technology (Banaras Hindu University) Varanasi - 221005, India}

\begin{abstract}
The Kitaev chain has been extensively explored in the context of uniform couplings, with studies focusing either on purely nearest-neighbor interactions or on systems dominated by long-range superconducting pairing. Building on these investigations, we introduce a hybrid Kitaev chain in which the lattice is partitioned into two segments: the left segment comprises nearest-neighbor couplings, while the right segment incorporates long-range pairing. To probe the role of the interface, we study two scenarios: a decoupled (suppressed hopping) case, where the segments are isolated, and a coupled case, where they are connected via interface hopping that enables tunneling. Using this setup, we investigate the behavior of Majorana zero modes at the interface between the two segments, finding that in the decoupled case, Majorana zero modes remain sharply localized at the left segment chain edges while massive Dirac modes remain in right segment chain edges, with their energies and localization strongly dependent on the long-range pairing exponent. Introducing a finite interface coupling enables transfer of Majorana zero modes from the edges of the left segment to those of the right segment of the chain. We characterize this dynamics by the fidelity of state transfer, dynamical rotation, and inverse participation ratio. We show the signature of Majorana zero mode transfer across the interface by the spatiotemporal profile of the probability distribution of the time evolved state. 
\end{abstract}
\maketitle

\section{Introduction}
   In recent years, topological systems in condensed matter such as insulators, superconductors and semi-metals have garnered significant attention \cite{RevModPhys.82.3045, RevModPhys.83.1057, RevModPhys.91.015005, PhysRevLett.100.096407, Sato_2017} particularly due to their characteristic topological protection \cite{aguado2017majorana}. A foundational model in this field is the one-dimensional topological superconductor introduced by Kitaev, commonly referred to as the  Kitaev chain \cite{A_Y_Kitaev-2001}. The excitations in this model are also their own antiparticles \cite{PhysRevLett.106.220402, PhysRevLett.121.076802}, which is an exotic property absent in conventional, non-topological superconductors. Importantly, the said excitations termed as the Majorana zero modes (MZMs) are topologically protected and remain resilient against local perturbations, disappearing only when the system undergoes a topological phase transition. This inherent robustness makes them highly promising candidates for fault-tolerant quantum computing due to non-Abelian statistics \cite{sarma2015majorana, flensberg2021engineered}  and manipulation in topological quantum computing architectures \cite{PhysRevLett.134.096601, PhysRevA.73.042313, PhysRevLett.120.230405, 10.1063/5.0102999, das2006topological, beenakker2013search, elliott2015colloquium, RevModPhys.80.1083, KITAEV20032, PhysRevLett.101.010501, PhysRevLett.94.166802}. Despite the theoretical elegance of the Kitaev chain, naturally occurring intrinsic \emph{p}-wave superconductors, which are necessary for realizing these topological phases because of their spin-triplet pairing and anisotropic gap structure, are extremely rare. To overcome this limitation, several theoretical proposals have demonstrated that effective \emph{p}-wave superconductivity can be engineered in one-dimensional systems using experimentally accessible platforms \cite{PhysRevB.61.10267,Dvir_2023, Bordin_2025}. These proposals aim to emulate the Kitaev chain physics in realistic settings. Examples include chains of magnetic adatoms \cite{PhysRevB.88.020407, science.1259327}, planar Josephson junctions \cite{PhysRevX.7.021032, PhysRevLett.118.107701, ren2019topological}, and proximitized semiconductor nanowires \cite{PhysRevLett.105.177002, PhysRevLett.105.077001, devices1995signatures, das2012zero}. In particular, semiconductor nanowires with strong spin-orbit coupling placed in proximity to conventional \emph{s}-wave superconductors, referred to as  semiconductor superconductor heterostructures \cite{PhysRevLett.104.040502, PhysRevB.82.214509,  Alicea_2012, PhysRevB.81.125318, PhysRevB.84.144522, PhysRevLett.124.096801, PhysRevB.98.155314}, have emerged as one of the most promising platforms for realizing MZMs. Subsequent active experimental efforts \cite{Dvir_2023, Bordin_2025, PhysRevB.109.035415, Deng_2016, Lutchyn_2018, r9pv-2prs, Mourik_2012, Zatelli_2024} in these nanowire systems have reported evidence consistent with MZMs \cite{li2018effect, PhysRevLett.105.077001, PhysRevB.82.214509, PhysRevLett.105.177002}, providing a concrete realization of Kitaev chain physics in realistic materials. However, the existence of MZMs in nanowire systems remains under active debate, but also several challenges like disorder \cite{Sau_2012, PhysRevB.103.224505}, impurity effects \cite{PhysRevB.110.214506, PhysRevApplied.16.054053}, and quasi-Majorana states \cite{PhysRevResearch.6.033314} can complicate the unambiguous identification of topological MZMs. Complementary to these nanowire platforms, recent experimental studies in coupled quantum-dot systems have reported signatures of MZMs at the so-called ``sweet spot'', where the hopping amplitude  matches the superconducting pairing \cite{Dvir_2023, Bordin_2025}. Under this condition, robust MZMs can be realized even in systems comprising only a few quantum dots, enabling controlled investigations of Majorana fusion and dynamics. Theoretically, similar phenomena have also been explored in Ref.~\cite{PhysRevResearch.6.033314, PhysRevB.111.104311}.

The Kitaev chain has been extensively generalized to include long-range (LR) inducing superconducting pairing \cite{PhysRevLett.113.156402, PhysRevB.94.125121, PhysRevA.93.032311, PhysRevB.97.155113, PhysRevB.96.125113, PhysRevB.88.165111, PhysRevLett.119.110601, Cai_2017, PhysRevA.97.062301, PhysRevB.111.104308, PhysRevB.111.155149, PhysRevB.110.064302, Maity_2020} in the presence of combined LR hopping and pairing terms \cite{Maity_2020, PhysRevLett.118.267002, PhysRevB.95.195160}. The strength of the interaction in a given pairs decay algebraically with distance as \(1/d_l^{\alpha}\). Here, \(d_l\) represents the distance between sites and exponent \(\alpha\) determines how quickly the interaction strength decreases with distance, with smaller \(\alpha\) corresponding to LR interactions. In LR Kitaev chain topological phases with \(\alpha < 1\),there also exist massive Dirac modes (MDMs) \cite{PhysRevLett.113.156402, PhysRevB.94.125121, PhysRevA.93.032311, PhysRevB.97.155113, PhysRevB.96.125113, PhysRevB.88.165111, PhysRevLett.119.110601, PhysRevA.97.062301, PhysRevB.111.104308, PhysRevB.111.155149, PhysRevB.110.064302}, which are protected by system symmetry, but are not robust in presence of arbitrary perturbations. These MDMs remain localized at the edges of the chain and are termed as subgap modes. Such emergent subgap edge quasi-particles are absent in the Kitaev chain topological phases with \(\alpha > 1\). There exist MZMs at edges, and this  highlights the essential ingredients governing the stability and emergence of topologically nontrivial phases. Contrary to the MZMs the MDMs wavefunction show finite overlap with the eigen states while MZMs wavefunction are well localized. Studies of finite-length Kitaev chain have explored spectral properties and the behavior of MZMs and MDMs \cite{PhysRevB.99.094523, Hegde_2015, Leumer_2020, PhysRevB.110.155436, PhysRevB.111.075415, dxb2-15jf, kxqp-57vj, kmpz-3ytk, v6ht-jq3l, 21fx-x962, PhysRevB.111.104308, PhysRevB.111.155410, PhysRevB.110.064302, PhysRevResearch.7.023183, lr2b-nmrk, n7bl-slgm, PhysRevB.103.045428, PhysRevResearch.3.013148, v7gb-5gq8, PhysRevB.107.L060304, kfwb-71wf, 69jq-rcsb, jzfv-ygmr, phx7-hs6k, 8drx-qktw, PhysRevB.107.035440, PhysRevB.106.184505, PhysRevB.110.L201111, PhysRevResearch.6.L012062, PhysRevB.110.125408, PhysRevB.110.224510, PhysRevB.111.024507, FU2021168564, PhysRevB.107.235146, PhysRevLett.134.030402, PhysRevB.100.184306, PhysRevB.100.014434, PhysRevResearch.6.033314,PhysRevB.111.104311}. Experimentally, several studies have reported evidence suggesting detection signatures of MZMs \cite{deng2012anomalous, PhysRevLett.110.126406,  PhysRevLett.110.217005, mourik2012signatures} and the realization of LR Kitaev chain \cite{PhysRevB.97.235114}. 


\par
In this work, we consider a special type of Kitaev chain in which the two halves of the chain exhibit different types of superconducting pairing interactions. Specifically, one segment of the chain features conventional nearest-neighbor (short-range) superconducting pairing, while the other segment incorporates LR superconducting pairing interactions with algebraically decaying strength. These two halves are connected via an interface coupling, forming what we refer to as a {\it hybrid Kitaev chain}. The motivation for studying this hybrid setup is twofold. First, it allows us to explore how the coexistence of short-range and LR interactions affects the emergence, localization, and hybridization of MZMs and MDMs. Second, such a model provides a platform for investigating tunable topological transitions and novel boundary phenomena that cannot be realized in a purely short-range or LR Kitaev chain. By systematically varying the interaction strengths and the interface coupling strength, this model offers a versatile framework for understanding the interplay between different interaction ranges in topological superconductors and their potential applications in quantum information processing.
\par 
This paper is organized as follows. In Sec.~\ref{Model}, we introduce the hybrid Kitaev chain and analyze its energy spectrum, along with the probability distribution of the MZMs. Sec.~\ref{Fidelity_section} presents the dynamical evolution of fidelity. In Sect.~\ref{MP_DR}, we discuss Majorana polarization and its dynamics under rotation. Sec.~\ref{PR} is devoted to the study of the inverse participation ratio (IPR). Finally, Sec.~\ref{Conclusion} concludes our findings.

\section{Model}
\label{Model}
\subsection{Kitaev chain}
\label{LKC}
We begin with the standard long-range Kitaev chain comprising $L$ sites with open boundary conditions, described by~\cite{A_Y_Kitaev-2001, Vodola_2016, PhysRevB.94.125121}:
\begin{eqnarray}
\hat H_{\rm LR} &=&
 - J \sum_{j=1}^{L-1} \left( c_j^\dagger c_{j+1} + \text{H.c.} \right)
 - \mu \sum_{j=1}^{L} \left( c_j^\dagger c_j - \frac{1}{2} \right) \nonumber \\
&& + \Delta \sum_{j=1}^{L-1} \sum_{l=1}^{L-j} \frac{1}{d_l^\alpha} \left( c_j c_{j+l} + \text{H.c.} \right),
\label{LKC}
\end{eqnarray}
where $c_j^\dagger$ ($c_j$) creates (annihilates) a spinless fermion at site $j$, obeying canonical anticommutation relations $\{c_l, c_j^\dagger\} = \delta_{l,j}$ and $\{c_l, c_j\} = \{c_l^\dagger, c_j^\dagger\} = 0$~\cite{A_Y_Kitaev-2001}. The parameters $\mu$, $J$, and $\Delta$ denote the chemical potential, hopping amplitude, and superconducting pairing strength, respectively. We will assume all paramters are real and $J$ = $\Delta$ = 1. The LR exponent $(\alpha>0)$ controls the spatial decay of the pairing strength: small exponent $\alpha<1$ corresponds to slowly decaying, genuinely long-range couplings, while large exponent $\alpha>1$ recovers the nearest-neighbour (NN) interaction. In the extreme limit $\alpha \to \infty$, the LR Kitaev chain reduces to the NN Kitaev chain~\cite{A_Y_Kitaev-2001,PhysRevLett.113.156402,Vodola_2016,PhysRevB.94.125121, PhysRevB.90.014507}:
\begin{eqnarray}
\hat H_{\rm NN} &=&
 - J \sum_{j=1}^{L-1} \left( c_j^\dagger c_{j+1} + \text{H.c.} \right)
 - \mu \sum_{j=1}^{L} \left( c_j^\dagger c_j - \frac{1}{2} \right) \nonumber \\
&& + \Delta \sum_{j=1}^{L-1} \left( c_j c_{j+1} + \text{H.c.} \right).
\label{KC}
\end{eqnarray}

\begin{figure}
\includegraphics[width=0.49\linewidth,height=0.40\linewidth]{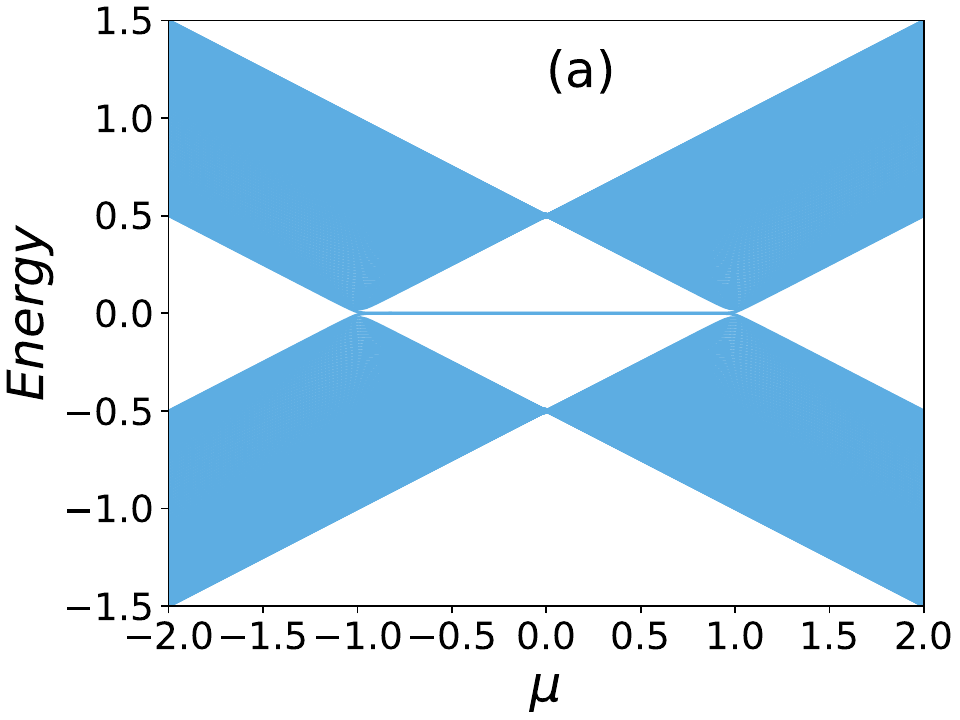}
\includegraphics[width=0.49\linewidth,height=0.40\linewidth]{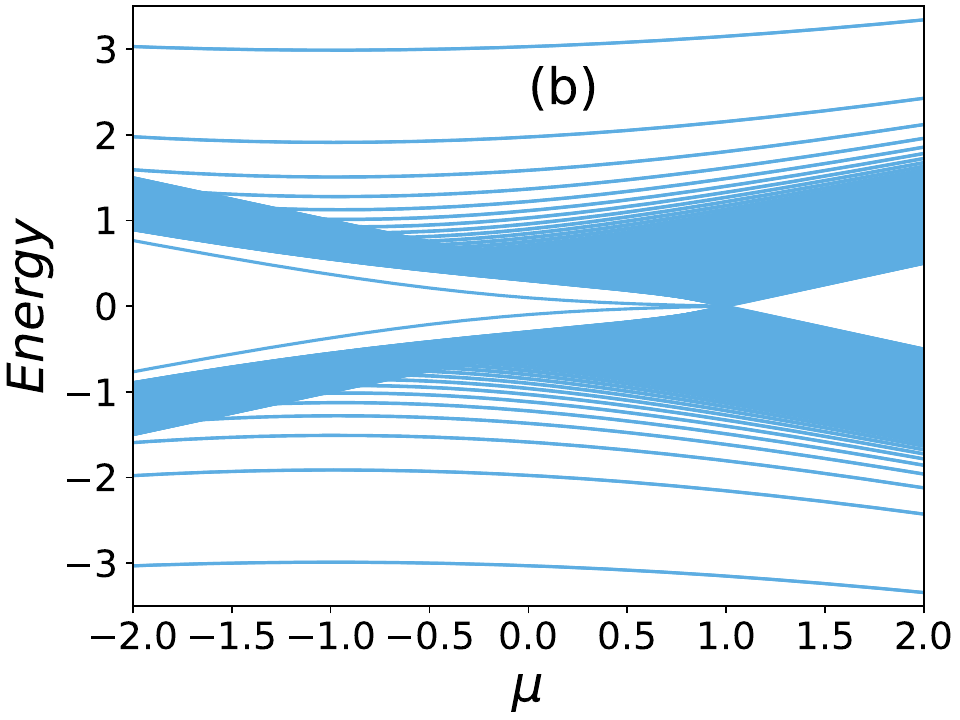}
    \caption{Energy spectrum of (a) a NN Kitaev chain and (b) a LR Kitaev chain under OBC, plotted as a function of the chemical potential $\mu$. The system parameters are $L = 100$ and $J = \Delta = 1.0$ and $\alpha = 0.5$ in LR Kitaev chain.}
  \label{ES_KC_LKC}
\end{figure}
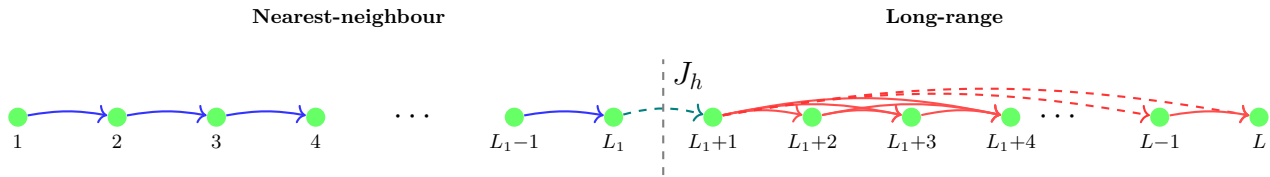
\begin{figure*}

\begin{tikzpicture}[scale=1.1, every node/.style={scale=0.85}]

\foreach \x [count=\i from 1] in {0,1.2,2.4,3.6} {
    \node[circle, fill=green!60, inner sep=3pt, label=below:\i] (s\i) at (\x, 2) {};
}

\node at (4.8,2) {\Large $\cdots$};

\node[circle, fill=green!60, inner sep=3pt, label=below:$L_1{-}1$] (s49) at (6.0, 2) {};
\node[circle, fill=green!60, inner sep=3pt, label=below:$L_1$] (s50) at (7.2, 2) {};

\draw[dashed, thick, gray] (7.8, 1.3) -- (7.8, 2.7);
\node at (4.0, 3.2) {\textbf{Nearest-neighbour}};
\node at (11.2, 3.2) {\textbf{Long-range}};

\foreach \x [count=\i from 1] in {8.4,9.6,10.8,12.0} {
    \pgfmathtruncatemacro{\idx}{\i + 50} 
    \node[circle, fill=green!60, inner sep=3pt, label=below:$L_1{+}\i$] (l\idx) at (\x, 2) {};
}

\node at (12.6,2) {\Large $\cdots$};

\node[circle, fill=green!60, inner sep=3pt, label=below:$L{-}1$] (l99) at (13.8, 2) {};
\node[circle, fill=green!60, inner sep=3pt, label=below:$L$] (l100) at (15.0, 2) {};

\foreach \i in {1,2,3} {
    \pgfmathtruncatemacro{\j}{\i + 1}
    \draw[->, thick, blue!80] (s\i) to[bend left=10] (s\j);
}
\draw[->, thick, blue!80] (s49) to[bend left=10] (s50);

\draw[->, thick, teal, dashed] (s50) to[bend left=15] (l51);
\node at (8.1, 2.5) {\Large $J_h$};

\foreach \i in {51,52,53,54} {
    \foreach \j in {51,52,53,54} {
        \ifnum\j>\i
            \draw[->, thick, red!70] (l\i) to[bend left=12] (l\j);
        \fi
    }
}

\foreach \i in {51} {
    \foreach \j in {99,100} {
        \draw[->, thick, red!80, dashed] (l\i) to[bend left=10] (l\j);
    }
}

\draw[->, thick, red!70] (l99) to[bend left=10] (l100); 
\end{tikzpicture}
\caption{{\bf A schematic diagram of a hybrid Kitaev chain:} The chain of even length $L$ is divided into two segments according to the interaction type: a NN  segment ($1 \leq j \leq L_1$) and a LR segment ($L_1 + 1 \leq j \leq L$). The two segments are connected by a interface coupling strength $J_h$  between sites $L_1$ and $L_1 + 1$. NN couplings are represented by blue arrows, while LR couplings are represented by red arrows.}
\label{FB}
\end{figure*}
Before discussing the  hybrid Kitaev chain, we first revisit the well-known energy spectra of the NN and LR Kitaev chain to establish a reference for comparison \cite{Maity_2020, Bhattacharya_2019, PhysRevB.105.085106}. The energy spectrum of the NN Kitaev chain [Fig.~\ref{ES_KC_LKC}(a)] exhibits symmetry under $\mu \leftrightarrow -\mu$, with topological phase transitions occurring at the two critical points $\mu = \pm J$. Within the range $-J < \mu < J$, the system is in a topological phase, hosting MZMs strictly localized at the chain edges. For $|\mu| > J$, the system enters a trivial phase with no edge modes.

The inclusion of LR superconducting pairing $\alpha$ brings about significant modifications to the energy spectrum, as evident from Fig.~\ref{ES_KC_LKC}(b). Notably, when $\alpha < 1$, the Kitaev chain departs from its NN Kitaev chain counterpart, with the critical point at $\mu = -J$ vanishing due to the predominance of LR interactions. The system remains trivial for $\mu > J$, whereas a topological phase emerges for $\mu < J$, now characterized by MDMs at the chain edges. These MDMs arise from the interaction of two MZMs at the both edges by LR pairing and appear at finite energy rather than zero energy. The $\mu \leftrightarrow -\mu$ spectral symmetry is broken for finite $\alpha < 1$, leaving only the transition at $\mu = J$. Furthermore, the energy gap and the localization of MDMs depend sensitively on both $\alpha$ and $\mu$, demonstrating tunable topological properties. The distinction between topological and trivial phases in NN and LR Kitaev chain establishes a natural baseline for the study of hybrid Kitaev chain. To this end, we consider two scenarios: one in which the interface coupling strength is absent (suppressed hopping) and another in which it is present. This comparison enables us to elucidate how the interplay between NN and LR interactions governs the structure of the energy spectrum and the emergence of edge modes across distinct phases.

\subsection{Hybrid Kitaev Chain}
\label{HKC}
We consider a chain of $L$ sites divided into two segments as shown in Fig.~(\ref{FB}), the first segment of length $L_1$ has NN interactions, and the second segment of length $L_2 (= L - L_1)$ has LR interactions. The interface connecting sites $L_1$ and $L_1+1$ allows coupling between the two segments with strength $J_h$. This NN–LR hybrid chain may be interpreted as a special limit of  LR–LR hybrid chain, obtained by taking the LR decay exponent of the left segment to infinity, effectively reducing it to a purely NN interactions.
\par
The coupled Hamiltonian of the hybrid system is expressed as the sum of three contributions: the NN interaction term ($\hat H^h_{\rm NN}$), the LR interaction term ($\hat H^h_{\rm LR}$), and the interface coupling term ($\hat H^h_{\rm I}$), mathematically given as 
\begin{equation}
\hat H^h = \hat H_{\rm NN}^h + \hat H_{\rm LR}^h + \hat H_{\rm I}^h,
\label{total_Ham_hybrid}
\end{equation}
with
\begin{eqnarray}
\hat H_{\rm NN}^h &=&
 - J \sum_{j=1}^{L_1-1} \left( c_j^\dagger c_{j+1} + \text{H.c.} \right)
 - \mu \sum_{j=1}^{L_1} \left( c_j^\dagger c_j - \frac{1}{2} \right) \nonumber \\
&& + \Delta \sum_{j=1}^{L_1-1} \left( c_j c_{j+1} + \text{H.c.} \right),
\label{H_NN_hybrid}
\end{eqnarray}

\begin{eqnarray}
\hat H_{\rm LR}^h &=& - J \sum_{j=L_1+1}^{L-1} \left( c_j^\dagger c_{j+1} + \text{H.c.} \right)
 - \mu \sum_{j=L_1+1}^{L} \left( c_j^\dagger c_j - \frac{1}{2} \right) \nonumber \\
&& + \Delta \sum_{j=L_1+1}^{L} \sum_{l=1}^{L-j} \frac{1}{d_l^\alpha} \left( c_j c_{j+l} + \text{H.c.} \right).
\label{H_LR_hybrid}
\end{eqnarray}

\begin{equation}
\hat H_{\rm I}^h =   J_h \left( c_{L_1}^\dagger c_{L_1+1} + c_{L_1}^\dagger c_{L_1+1}^\dagger + \text{H.c.} \right),
\label{H_Interface_hybrid}
\end{equation}

Here, $J_h$ denotes the interface coupling strength that simultaneously governs both electron hopping and superconducting pairing respectively. 

In general, the interface Hamiltonian can be written as
\begin{equation}
\hat{H}_{\mathrm{I}}^h = 
J_h \left( c_{L_1}^\dagger c_{L_1+1} + \mathrm{H.c.} \right)
+ \Delta_h \left( c_{L_1}^\dagger c_{L_1+1}^\dagger + \mathrm{H.c.} \right),
\label{H_Interface_hybrid_general}
\end{equation}
where $J_h$ and $\Delta_h$ describe the hopping and pairing amplitudes at the interface, respectively. The form used in Eq.~\eqref{H_Interface_hybrid}, which we adopt throughout this work, corresponds to the symmetric case $(J_h = \Delta_h)$ of Eq.~\eqref{H_Interface_hybrid_general}. 
As a result, the $p$-wave nature of the superconducting pairing is preserved across the interface, maintaining the \emph{topological protection and spatial localization} of the MZMs during their hybridization and transportation.

\subsubsection{Energy spectrum without interface coupling strength}

\begin{figure}
\includegraphics[width=0.49\linewidth,height=0.40\linewidth]{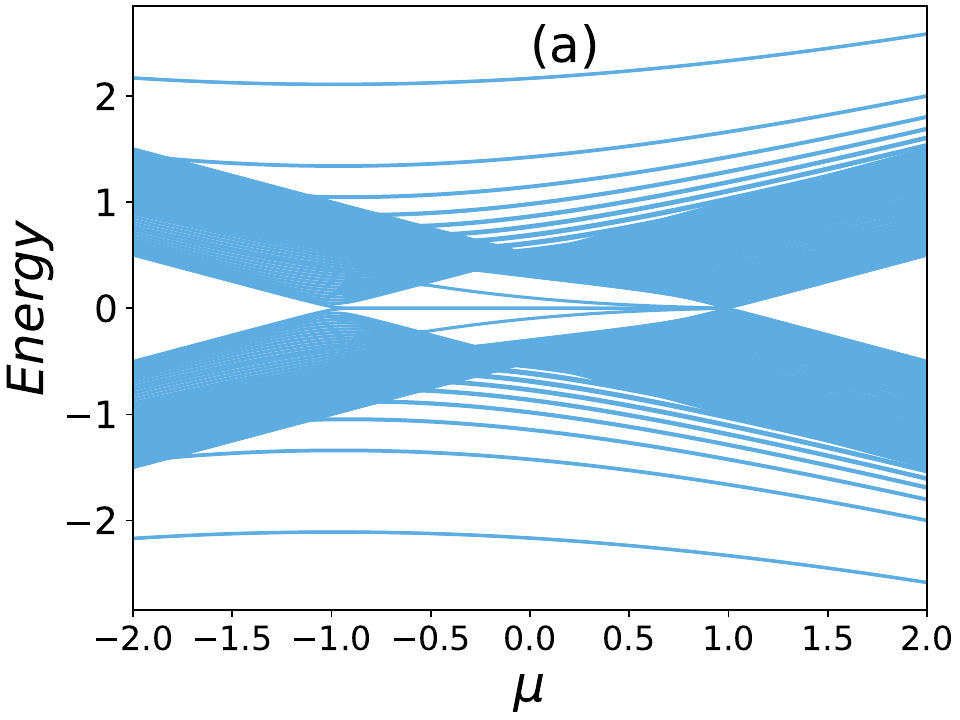}
\includegraphics[width=0.49\linewidth,height=0.40\linewidth]{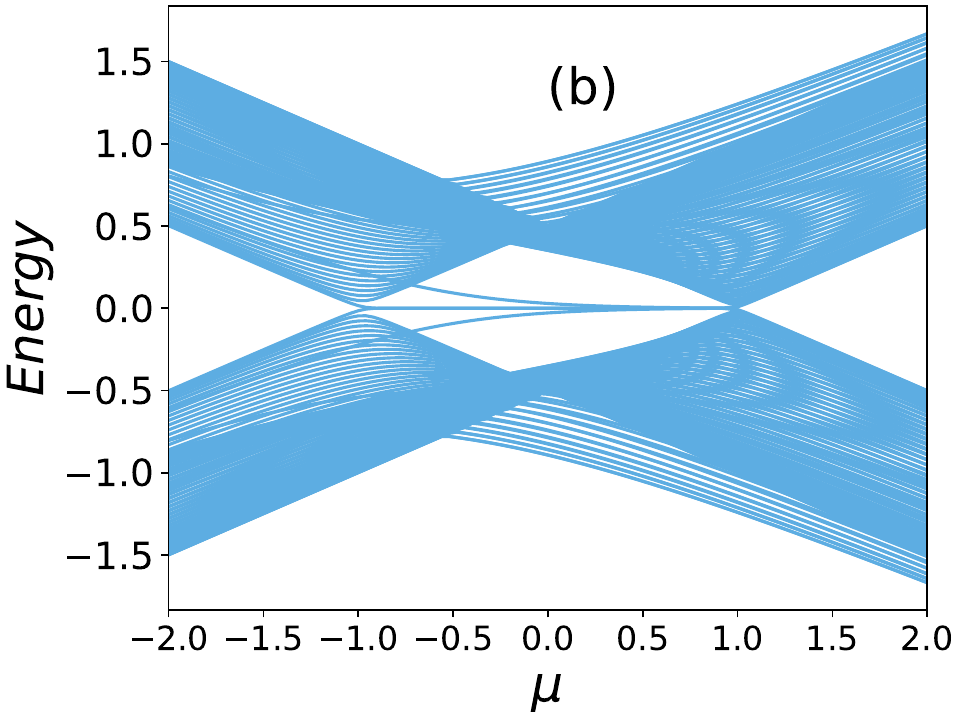}
\caption{{\bf Interface coupling strength absent ($J_h = 0.0$):} Energy spectrum of a hybrid Kitaev chain composed of a NN segment ($1 \leq j \leq L_1$) and an LR segment ($L_1 + 1 \leq j \leq L$), with $L_1 = 50$, total system size $L = 100$, under OBC. The system parameters are $J = \Delta = 1.0$. The spectrum is shown as a function of the chemical potential $\mu$ for decay exponents: (a) $\alpha = 0.5$ and (b) $\alpha = 1.0$.}
\label{Energy J_h = 0.0}
\end{figure}

Initially, starting with decoupled chains, ($J_h = 0$): the hybrid Kitaev chain effectively behaves as two independent chains: a NN segment and a LR segment, serving as instances of the corresponding Kitaev chain. The energy spectrum, shown in Figs.~\ref{Energy J_h = 0.0}(a,b), exhibits independent topological transitions for each segment. Due to the NN segment, MZMs appear in the topological regime $-J < \mu < J$, while resulting from the LR segment, MDMs emerge in the topological regime $\mu < -J$ . The two transition points at $\mu = -J$ and $\mu = +J$ highlight the differing pairing structures of the two segments. Increasing the LR decay exponent $\alpha$ reduces the MDMs bulk gap and enhances hybridization with MZMs. In the extreme LR exponent ($\alpha \gg 1$), both independent segments behave as NN Kitaev chains, producing four MZMs: two MZMs from the NN segment and two MZMs hybridized modes from the MDMs corresponding LR segment.

\subsubsection{Energy spectrum with interface coupling strength}
\begin{figure}
\includegraphics[width=0.49\linewidth,height=0.40\linewidth]{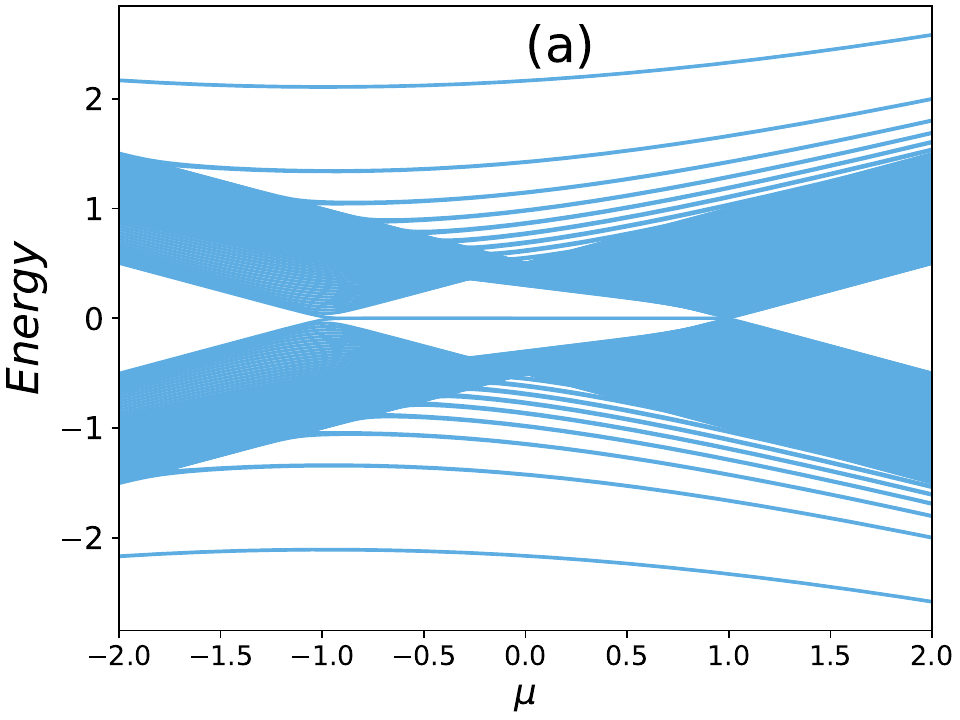}
\includegraphics[width=0.49\linewidth,height=0.40\linewidth]{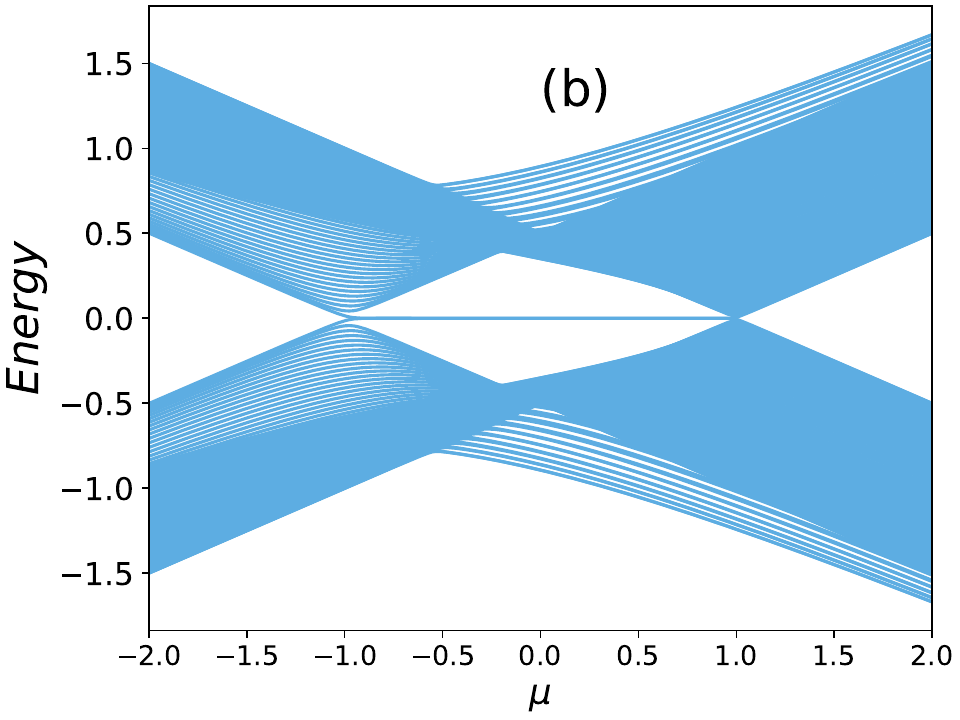}
\caption{{\bf Interface coupling strength present ($J_h = 1.0$):} Energy spectrum of a hybrid Kitaev chain composed of a NN segment ($1 \leq j \leq L_1$) and an LR segment ($L_1 + 1 \leq i \leq L$), with $L_1 = 50$, total system size $L = 100$, and OBC. The system parameters are $J = \Delta = 1.0$. The spectrum is shown as a function of the chemical potential $\mu$ for decay exponents: (a) $\alpha = 0.5$ and (b) $\alpha = 1.0$.}
\label{Energy J_h = 1.0}
\end{figure}

Introducing interface coupling strength ($J_h = 1.0$) qualitatively alters the spectrum. The finite-energy MDMs associated with the LR segment are suppressed, while zero-energy MZMs remain robust, as shown in Fig.~\ref{Energy J_h = 1.0}(a,b). For small $\alpha <1$ (strongly long-range), the topological regime $|\mu| < J$ supports two MZMs accompanied by additional finite-energy states originating from LR pairing [Fig.~\ref{Energy J_h = 1.0}(a)]. For intermediate $\alpha$ ($\alpha \sim 1$), the spectrum resembles the NN Kitaev chain: only two zero-energy MZMs persist, while finite-energy LR modes are suppressed [Figs.~\ref{Energy J_h = 1.0}(b)]. 

These results demonstrate that the interface coupling plays a key role: it stabilizes the zero-energy MZMs while suppressing the finite-energy modes induced by LR interactions. The hybrid Kitaev chain therefore provides a continuous interpolation between the behavior of a conventional NN Kitaev chain and that of a LR Kitaev chain, with the parameter $\alpha$ controlling the strength of LR effects. Its energy spectrum clearly reflects the competition between NN and LR pairing. A direct comparison with the pure NN Kitaev chain shows that LR correlations and hybridization give rise to new edge phenomena, whereas the presence of interface coupling helps recover the familiar topological features of the standard Kitaev chain.
\par
We analyze the energy spectrum of the LR–LR hybrid Kitaev chain by fixing the interface coupling strength $J_h$, in which both segments feature LR interactions with potentially different decay exponents $(\beta, \alpha)$ (see Appendix~\ref{LR_LR_hybrid} for details). This LR–LR setup generalizes the NN–LR model and allows us to investigate how variations in LR interactions and the interface coupling influence the low-energy spectrum and the localization of MZMs. As shown in Appendix~\ref{LR_LR_hybrid}, for weak interface coupling ($J_h \to 0$), the system behaves as two independent LR interacting chains, hosting four MDMs (two per segment) localized at their segment edges. For strong interface coupling ($J_h \to 1$), the segments merge into a single LR chain, reducing the number of MDMs to two, localized at the edges of the full chain, while the finite-energy subgap states redistribute across the spectrum. Despite these changes in the low-energy structure, the MDMs remain robust, and the topological transitions are preserved. Representative spectra, along with a detailed discussion of the Hamiltonian and interface effects, are provided in Appendix~\ref{LR_LR_hybrid}.

For completeness, we also consider the more general case of unequal interface hopping and pairing amplitudes ($J_h \neq \Delta_h$) for the set up NN–LR hybrid chain. As detailed discussed in Appendix~\ref{Interface_coupling}, such asymmetry primarily modifies the low-energy structure without affecting the topological properties of the chain. For moderate detuning, the spectrum remains largely unchanged, closely resembling the symmetric case apart from small shifts in avoided crossings due to altered hybridization between the NN and LR segments. Even in strongly imbalanced regimes, where the pairing term pulls additional finite-energy states into the gap, the topological phase boundaries and the existence of both MZMs and MDMs remain intact. Representative spectra for different $(J_h, \Delta_h)$ values are provided in Appendix~\ref{Interface_coupling}.

\begin{figure}
\includegraphics[width=0.49\linewidth,height=0.40\linewidth]{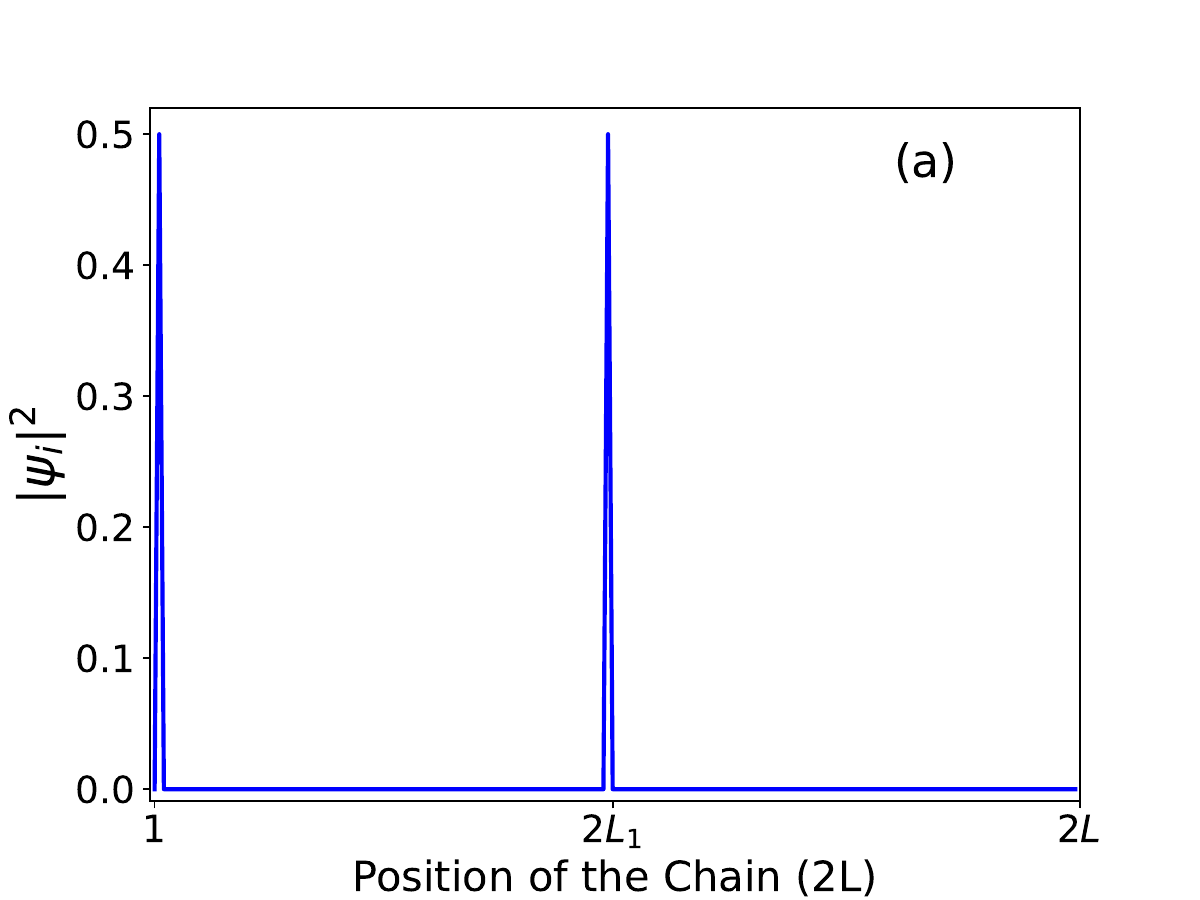}
\includegraphics[width=0.49\linewidth,height=0.40\linewidth]{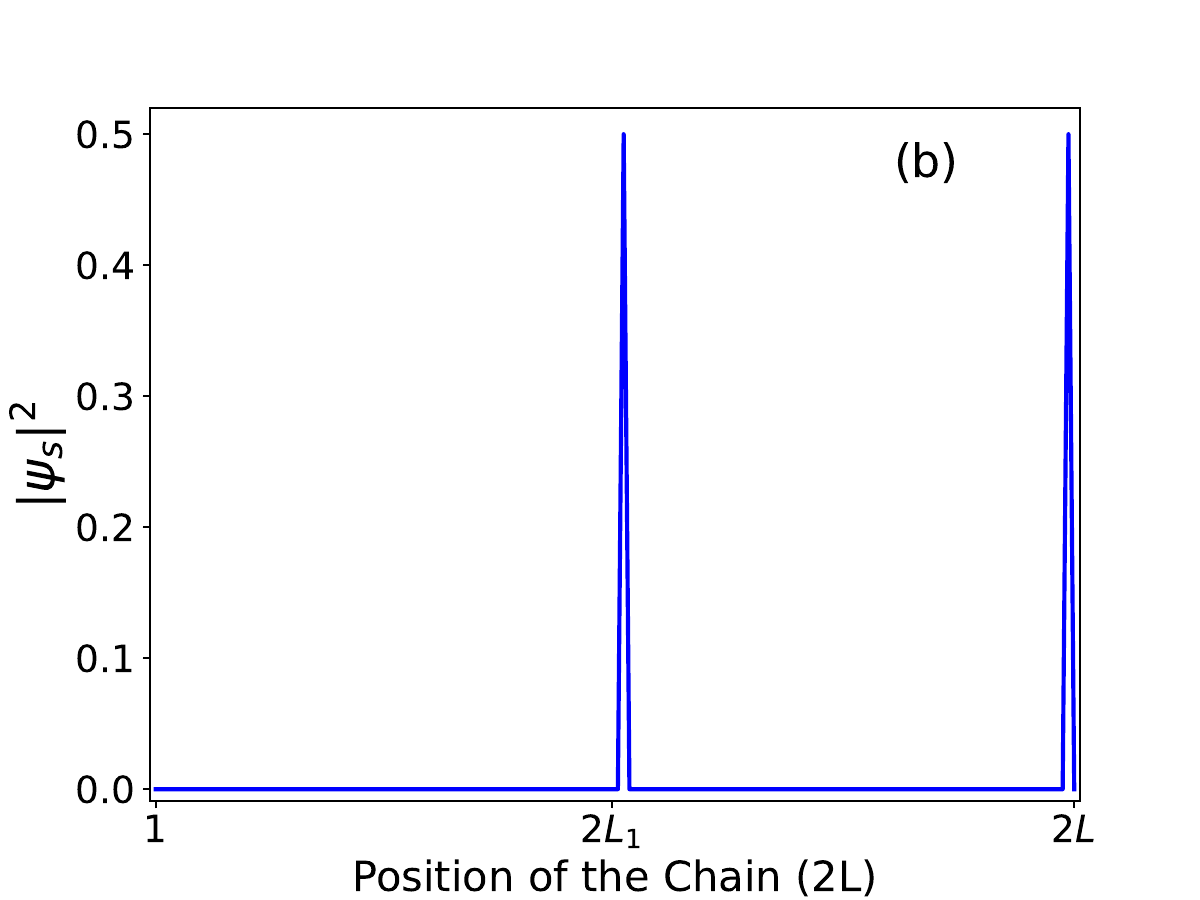}
\caption{Probability distribution of MZMs along the sites of a hybrid Kitaev chain of length \(L\), where \(2L\) represents the total length spanned by the Majoranas, under open boundary conditions. Panel (a) shows the nearest-neighbor (NN) interacting chain on the left and the long-range chain on the right, while panel (b) shows the switched configuration, if the MZM were localized on the edges of the right segment. The system parameters are \(J = \Delta = 1.0\), \(\mu = 0.0\), \(J_h = 0.0\), and \(\alpha = 0.5\).}
\label{MZMs}
\end{figure}

\subsection{Probability Distribution of MZMs}
\label{SPD_MZM}

To investigate the spatial structure of MZMs in the hybrid Kitaev chain, we first consider the decoupled limit with interface coupling set to $J_h = 0$ (suppressed hopping). In this configuration, both  segments of the chain remain independent. The left segment have NN Kitaev chain, they hosts MZMs strictly localized at its edges [Fig.~\ref{MZMs}(a)]. The right segment have LR Kitaev chain, that support  MDMs, depending on the decay exponent $\alpha$, but the primary focus here is on the zero-energy modes. For this purpose, we switch the Hamiltonian configuration, i.e. left segment of the hybrid chain have LR interaction and right segment of hybrid chain which has NN interaction, in the decoupled limit, the  MZMs relocates to the edges of right segment [Fig.~\ref{MZMs}(b)], reflecting a switching of the system’s topological character. This setup serves as a reference switch state to study MZMs transport from the left segment to the right segment of the chain, if driven purely by topological changes.
\par
In the limit of $J_h\rightarrow 0$,  the hybrid Kitaev chain effectively behaves as two independent chains. The hybrid chain in their topological regions hosts four zero-energy modes: two conventional MZMs localized at the edges of the NN Kitaev chain segment and two additional modes emerging from the LR Kitaev chain segment, with LR exponent $\alpha>1.0$. Here, finite-energy MDMs collapse to true zero-energy modes. In contrast, for strong LR interactions ($\alpha < 1.0$), the LR Kitaev chain segment supports two finite-energy MDMs, and only the two MZMs remain at the edges of the left segment.
\par
This analysis demonstrates that the  energy spectrum, and spatial localization of MZMs in the hybrid Kitaev chain can be continuously tuned by varying the LR exponent and interface coupling strength. The interplay between NN and LR interactions provides a rich topological landscape, where MZMs can be transported, hybridized, or fully localized, highlighting the controllable nature of edge states in hybrid topological systems.

\section{Dynamical response of Majorana Zero modes}
\label{Fidelity_section}
To analyze the dynamical response of the MZMs to a sudden quench of the interaction profile, we compute the fidelity between the time evolved initial state and a fixed target (switch) state. The system is initially prepared in an eigenstate $|\psi_i\rangle$ of the decoupled hybrid Hamiltonian $\hat{H}^h$, which hosts MZMs at left segment edges. Following a quench, the Hamiltonian configuration is switched so that the left segment of the chain hosts LR interactions while the right segment hosts NN interactions. For this new switch Hamiltonian, the MZMs relocate to the edges of the right segment. Consequently, the initial MZM is no longer an eigenstate and starts to hybridize with the bulk modes of the post-quench configuration. The corresponding target state $|\psi_s\rangle$ is defined as the ideal MZM eigenstate of $\hat{H}^h_S$, representing a perfectly localized mode in the new setup. Since both the initial and quenched configurations host two zero-energy modes at the chain edges, we represent the initial state $|\psi_i\rangle$ and the target state $|\psi_s\rangle$ as superpositions of the corresponding zero-energy modes.

The dynamical response of the MZMs to the quench is then quantified using the fidelity \cite{PhysRevResearch.3.013148,laflorencie2023universal, PhysRevB.85.024301, PhysRevE.74.031123, PhysRevLett.99.095701, PhysRevLett.100.080601, PhysRevB.80.014403, PhysRevLett.103.170501, PhysRevB.83.075118, PhysRevLett.106.055701, PhysRevA.79.060301,ye2025controlling} defined as 
\begin{equation}
\label{fidelity}
\mathcal{F}(t) = |\langle\psi_s|\psi_i(t)\rangle|^2,\quad \ket{\psi_i(t)} = e^{-i \hat{H}^h t} \ket{\psi_i}. 
\end{equation}
 This quantity measures the overlap between the time-evolved initial state and the target MZM eigenstates, serving as a direct probe of the topological edge state's stability. The ensuing temporal evolution of the fidelity reveals the nature of the system's adaptation: a value persistently near unity signifies a robust mode that remains localized under the new Hamiltonian, indicating strong topological protection. Conversely, a decay towards zero marks the fragility of the initial state and its dissolution into the bulk, with the decay rate quantifying the speed of information loss. Should the fidelity exhibit sustained oscillations, it signifies that the initial MZM has hybridized with another state—such as the MZM at the opposite edge to form a finite-energy fermion; in this case, the oscillation frequency provides a direct measure of the energy splitting inherent to this hybridization, a critical metric for assessing the viability of topological qubit operations.
\par

\begin{figure}
    \centering
    \includegraphics[width = 0.49\linewidth,height=0.40\linewidth]{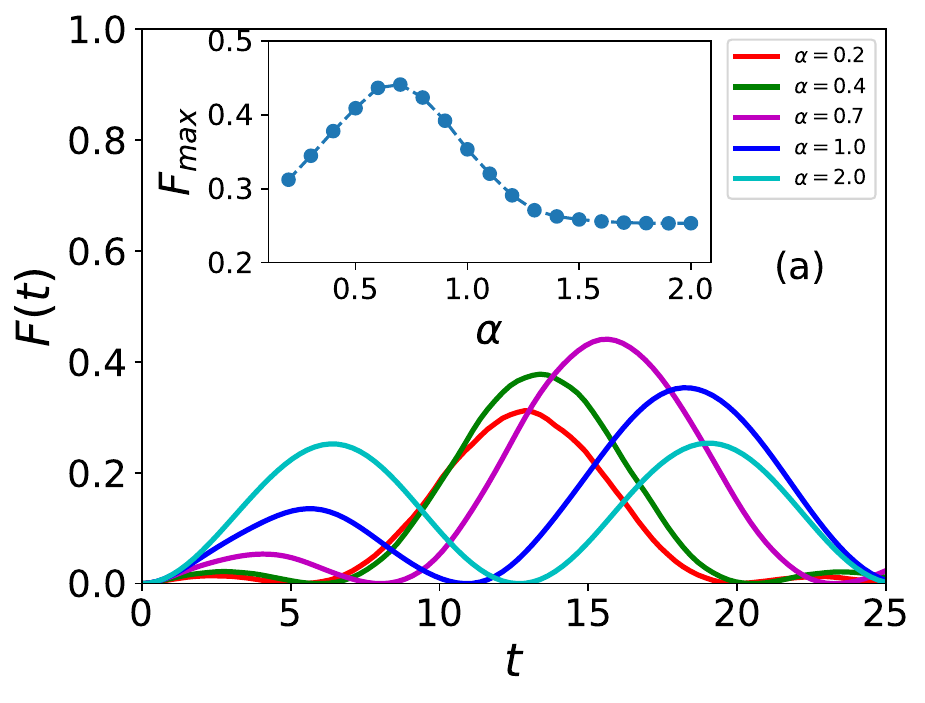}
    \includegraphics[width = 0.49\linewidth,height=0.40\linewidth]{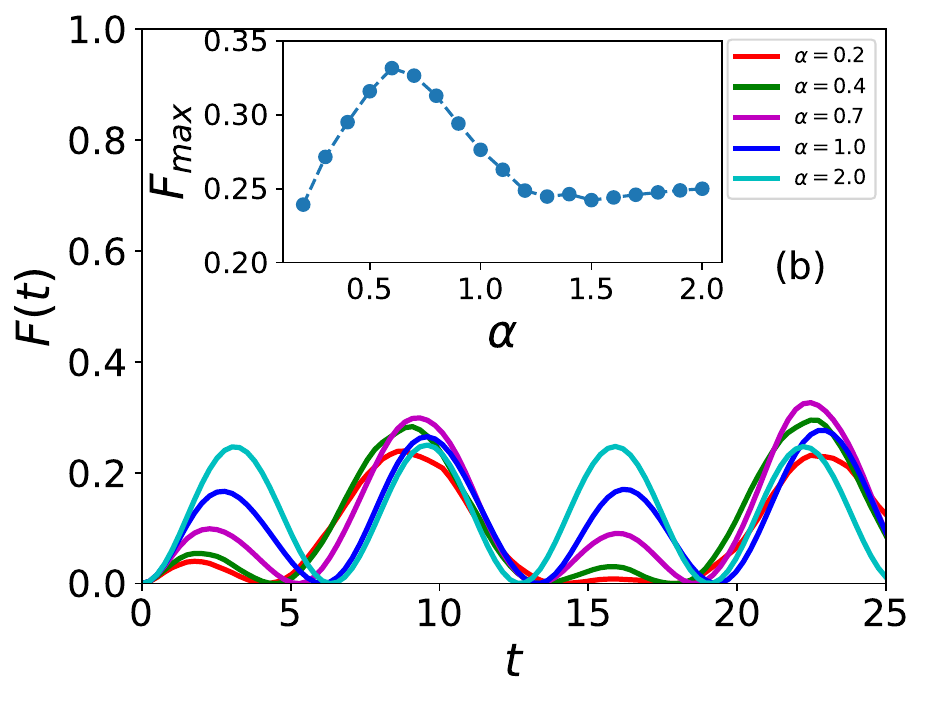}
\caption{Fidelity as a function of time for different decay exponents \(\alpha\) (indicated in the legends) with system size \(L = 100\) and interface coupling strengths: (a) \(J_h = 0.5\) and (b) \(J_h = 1.0\). The insets show the corresponding maxima of the fidelity as a function of the decay exponent \(\alpha\).}
\label{Fidelity}
\end{figure}

We analyze the fidelity dynamics of the hybrid Kitaev chain to quantify the transfer of MZMs across the interface between the left and right  segments. First, we consider the decoupled case ($J_h = 0.0$), where the fidelity remains zero, namely, $|\langle\psi_s|\psi_i(t)\rangle|^2 = 0$ at all times, because the left segment and the  right segment are fully independent. Therefore, in the absence of coupling across the interface, no transfer of MZMs occurs, and the system exhibits no nontrivial dynamics. This clearly demonstrates that interface (inter-segment) coupling strength is an essential requirement for the dynamical evolution of the MZM state.

Next, we switch on the interface coupling with a weak strength ($J_h = 0.5$) and vary the LR exponent $\alpha$ in the range $0.2 \leq \alpha \leq 2.0$. In this regime, the fidelity exhibits  oscillations whose amplitude depends sensitively on $\alpha$ [Fig.\ref{Fidelity}(a)]. The maximum fidelity increases with $\alpha$, reaching its peak at $\alpha = 0.7$, and decreases for larger $\alpha$. This enhancement arises from improved spatial localization and greater energetic separation of the MZMs at intermediate $\alpha$, which facilitates tunneling across the weakly coupled interface. The weak coupling also acts as a protective barrier, suppressing hybridization with bulk states and preserving fidelity [inset of Fig.\ref{Fidelity}(a)]. For the LR exponent $(\alpha \geq 0.7)$, fidelity declines with increasing $\alpha$. For $J_h = 0.5$, $F_{\rm max}$ decreases from $\approx 0.44$ at $\alpha = 0.7$ to $\approx 0.25$ at $\alpha = 2.0$, as the LR interaction segment approaches the NN interaction limit and MZMs become more localized, reducing tunneling across the interface. Upon increasing the interface coupling to normal coupling $J_h = 1.0$, the fidelity maxima shows similar behavior as $J_h=0.5$ [Fig.\ref{Fidelity}(b)]. The maximum fidelity again increases with $\alpha$, peaking at $\alpha = 0.7$ before decreasing at higher values. This trend reflects a competition between improved MZM localization and reduced interface overlap due to excessive localization at larger $\alpha$ [inset of Fig.\ref{Fidelity}(b)].

A key observation is the maximum fidelity is larger for weak coupling $J_h = 0.5$ than for normal coupling $J_h = 1.0$ at a LR exponent $\alpha = 0.7$. This indicates that the MZM transfer is more localized for weaker interface coupling compared to normal coupling. This counterintuitive result arises from the hybrid Kitaev chain inherent topological mismatch: strong coupling enhances hybridization between MZMs and bulk states near the interface, leading to destructive interference and reduced transfer efficiency. Weak coupling, in contrast, acts as a controlled tunnel barrier, enabling  MZM transfer while minimizing bulk mixing.

\section{Majorana Polarization and Dynamical Rotation}
\label{MP_DR}

MZMs in a Kitaev chain or related systems are characterized by their particle-hole symmetric nature and spatial localization. A key quantity for identifying and analyzing these modes is the \emph{Majorana polarization}  \cite{PhysRevB.110.165404, Sedlmayr_2015, PhysRevB.92.115115, PhysRevLett.108.096802, PhysRevB.102.085411, PhysRevLett.108.096802, PhysRevB.78.195125}, which quantifies the degree to which a state exhibits localization character. Specifically, Majorana polarization captures how the electron-like and hole-like components are distributed along the chain and provides a spatially resolved map of Majorana localization. A value of Majorana polarization close to unity indicates a state with strong Majorana nature, whereas values near zero correspond to ordinary bulk fermionic states. Evaluating Majorana polarization locally is particularly useful for detecting hybridization or delocalization of MZMs in inhomogeneous hybrid Kitaev chain.

In the spinless fermionic Kitaev chain, particle-hole symmetry is encoded in the antiunitary operator

\begin{equation}
\mathcal{C} = \tau^x \otimes \mathbb{I}_L \, \mathcal{K},
\end{equation}
where \(\tau^x\) is the Pauli matrix acting in particle-hole space, \(\mathbb{I}_L\) is the identity operator in the lattice site space, and \(\mathcal{K}\) denotes complex conjugation. This operator satisfies \(\mathcal{C}^2 = 1\), ensuring that its eigenstates can be classified according to particle-hole symmetry. Here, \(\ket{\gamma}\) represents an eigenstate of the system, which may be either a MZM or a conventional bulk excitation. The Majorana polarization of MZM state \(\ket{\gamma}\) is then defined as
\begin{equation}
R =  \braket{\gamma | \mathcal{C} | \gamma},
\end{equation}
where \(R = -1\) identifies a perfectly localized MZM at edges of chain, while \(R \approx 0\) corresponds  for finite energy states. Evaluating \(R\) locally along the chain provides a spatially resolved map of the Majorana character, indicating where the MZMs are localized and how strongly they hybridize with the bulk.

To study the time evolution of MZMs, we consider an initial state $\ket{\psi_{i}}$, typically an eigenstate of the decoupled hybrid Hamiltonian, and evolve it under the hybrid configuration of Hamiltonian $\hat H^h$ :
\begin{equation}
\ket{\psi(t)} = e^{-i \hat H^h t} \ket{\psi_{i}}.
\end{equation}
The \emph{dynamical rotation} probes how the particle-hole symmetry of this state evolves over time. This is formalized by introducing a particle-hole rotation operator
\begin{equation}
U_{\rm PH}(\theta) = e^{-i \theta \mathcal{C}},
\end{equation}
where $\mathcal{C} = \tau^x \otimes \mathbb{I}_L $. For $\theta = 1$, the dynamical overlap with the target MZM state $\ket{\psi_{s}}$ (MZM of the switched Hamiltonian configuration) of the Hamiltonian is defined as
\begin{equation}
R(t) = \braket{\psi_{s} | U_{\rm PH}(1) | \psi(t)}.
\end{equation}
In this convention, a perfectly localized MZM corresponds to a $\mathcal{R}$-eigenvalue of $-1$. When $\ket{\psi(t)}$ remains in this eigenspace, $R(t)$ reaches its most negative real value, indicating maximal overlap with a particle-hole symmetric localized MZM. Deviations from this value indicate hybridization, delocalization, or loss of MZM localization character.

\begin{figure}
    \centering
    \includegraphics[width = 0.49\linewidth,height=0.40\linewidth]{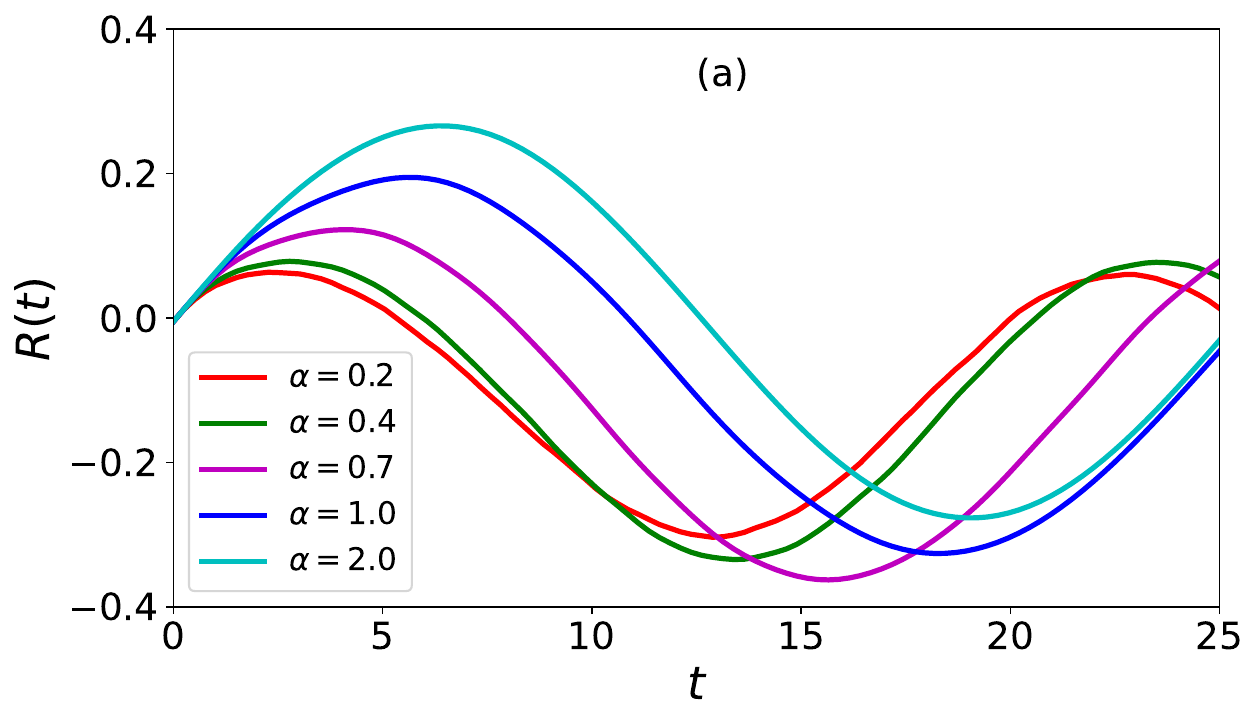}
    \includegraphics[width = 0.49\linewidth,height=0.40\linewidth]{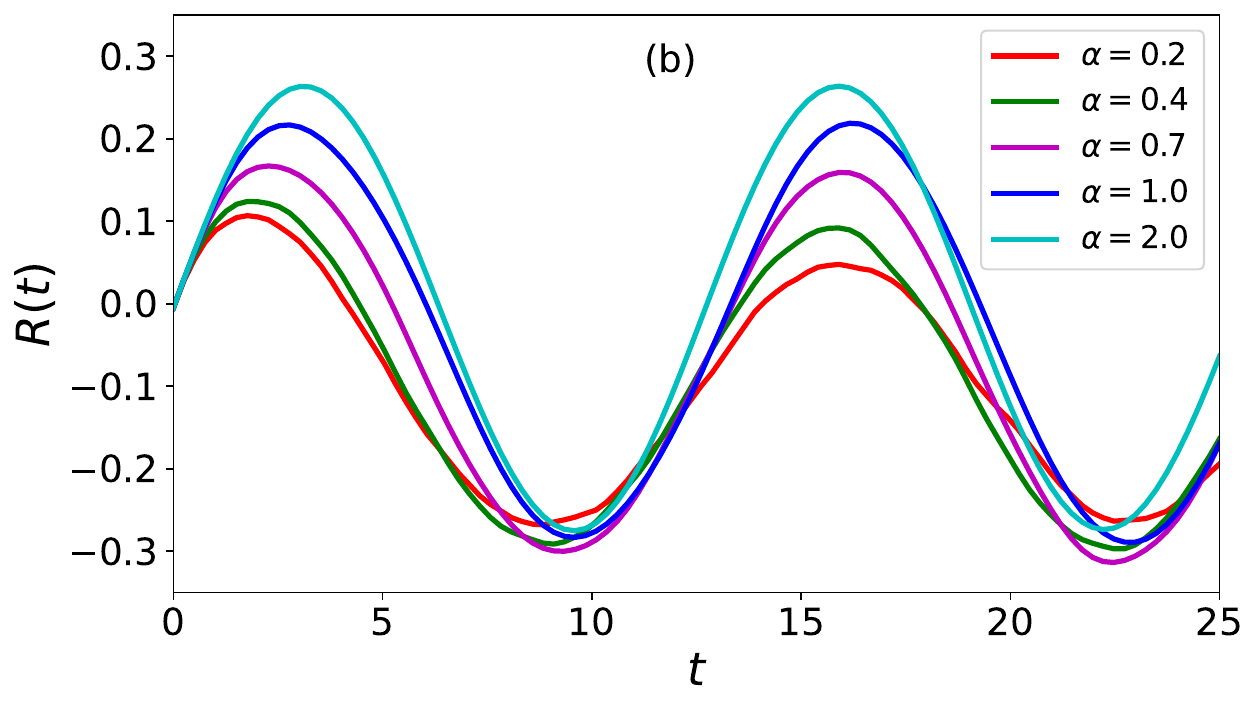}
  \caption{Time evolution of particle-hole rotation for various decay exponents \(\alpha\) (indicated in the legends) with system size \(L = 100\), presented for two interface coupling strengths: (a) \(J_h = 0.5\) and (b) \(J_h = 1.0\).}
  \label{MP}
\end{figure}

We analyze the dynamical evolution of particle-hole rotation under the hybrid Hamiltonian $\hat H^h$ to probe the time-dependent behavior of MZMs and their overlap with the target (switched) state. Fig.~\ref{MP} show the particle-hole rotation for different LR exponents $0.2 \leq \alpha \leq 2.0$. For each $\alpha$, the dynamics is computed for two representative interface coupling strength: weak coupling strength ($J_h = 0.5$) and normal coupling strength ($J_h = 1.0$).

For weak interface coupling ($J_h = 0.5$), the particle-hole rotation exhibits regular, high-amplitude oscillations, particularly in the strong LR regime ($\alpha < 1.0$) [Fig.~\ref{MP}(a)]. This regularity mirrors the higher fidelity in this regime, demonstrating that weak coupling reduces hybridization with bulk states during MZM transfer. Conversely, for stronger coupling ($J_h = 1.0$), the particle-hole rotation dynamics become more irregular and damped for small $\alpha$, consistent with reduced fidelity due to increased interference from bulk modes near the interface [Fig.~\ref{MP}(b)].

The rate and amplitude of particle-hole rotation are strongly dependent on $\alpha$. As $\alpha$ increases, the oscillation period lengthens and the amplitude diminishes, reflecting enhanced localization of MZMs and reduced overlap across the interface. This trend provides a clear dynamical signature of the transition from LR to short-ranged pairing and complements the static topological characterization obtained from energy spectra and Majorana polarization.

These results demonstrate that MZM transfer in the hybrid Kitaev chain is controlled by both the LR exponent $\alpha$ and the interface coupling $J_h$, offering a time resolved perspective on topological state evolution and the interplay between localization and hybridization.

\begin{figure}
    \centering
    \includegraphics[width = 0.48\linewidth,height=0.40\linewidth]{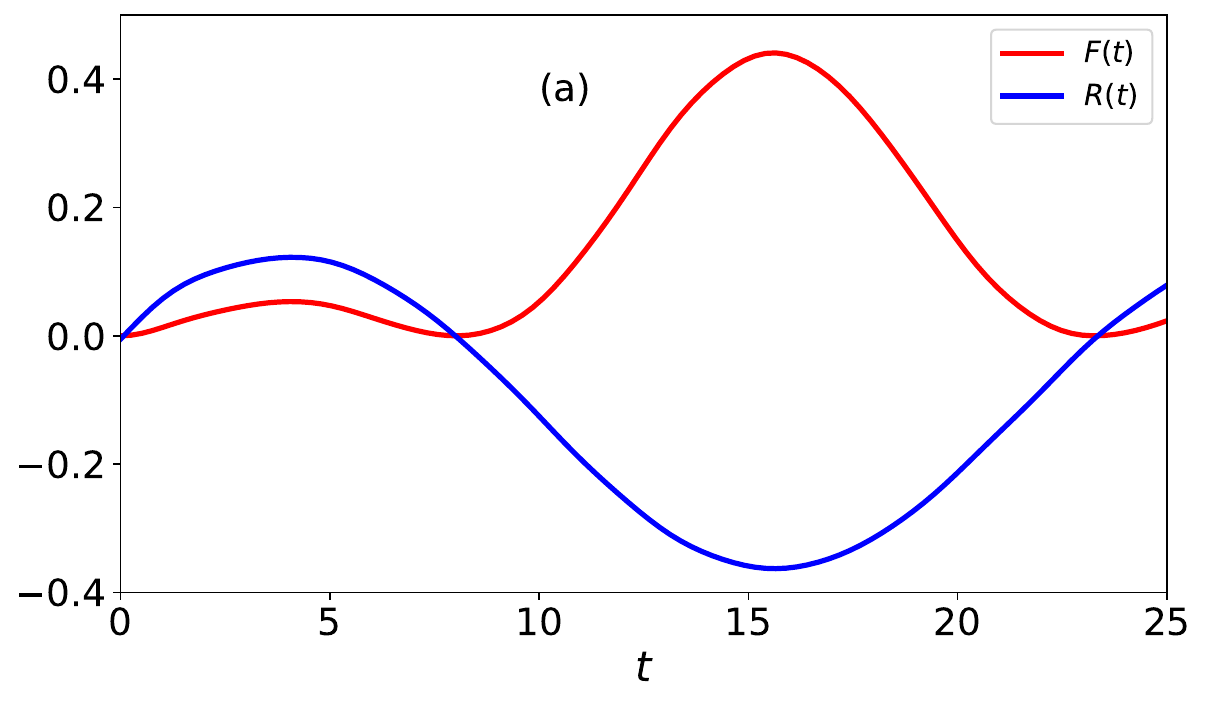}
    \includegraphics[width = 0.48\linewidth,height=0.40\linewidth]{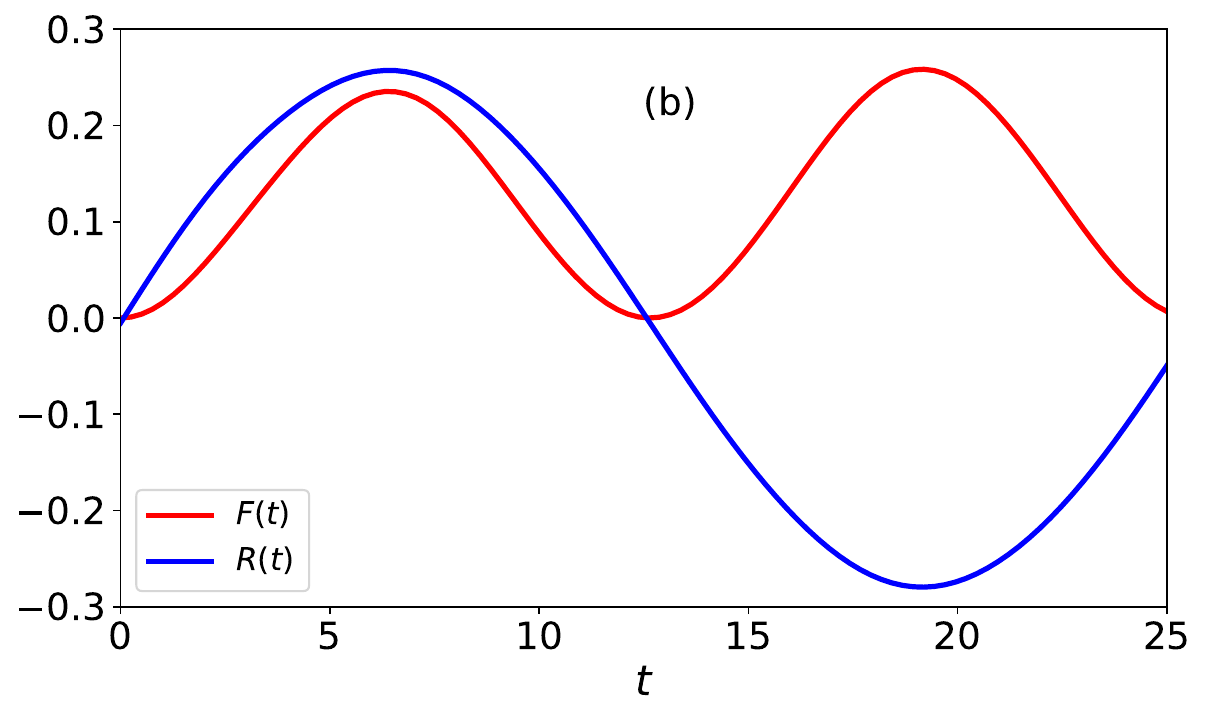}
\caption{Time evolution of fidelity (red) and dynamical rotation (blue) for interface coupling strength \(J_h = 0.7\), shown for decay exponents: (a) \(\alpha = 0.7\) and (b) \(\alpha = 1.5\) with system size \(N = 100\).}
\label{MP_Fidelity}
\end{figure}

{\it Correlation between Fidelity and Dynamical Rotation:} There is a correlation between fidelity and the dynamical rotation in the hybrid Kitaev chain. Maxima in fidelity coincide with the most negative values of dynamical rotation [Fig~\ref{MP_Fidelity}], indicating that maximal overlap with the switched reference state occurs when the system is fully localized in the $R=-1$ particle-hole symmetric sector. Deviations of dynamical rotation from this extremum correspond to partial delocalization or departure from the particle-hole symmetric subspace, while values near zero indicate orthogonality to the particle-hole localized reference state. This establishes dynamical rotation as a symmetry-resolved probe that complements fidelity in characterizing MZM dynamics.

In the LR interacting case ($\alpha = 0.7$), the extended nature of the pairing enhances the spatial overlap of MZMs across the interface, enabling partial transfer between segments. The alignment of fidelity maxima with the negative peaks of dynamical rotation confirms that  MZM transfer is accompanied by strong particle-hole symmetry localization. The slower oscillation frequency reflects the weaker effective coupling between MZMs in the LR regime [Fig~\ref{MP_Fidelity} (a)].

For larger LR exponents ($\alpha = 1.5$), portion of LR interacting system approaches  NN interacting case and system behaves effectively as a unified NN Kitaev chain. The preexisting edge MZMs lead to higher-frequency fidelity oscillations, indicating stronger Majorana coupling. While the correlation between fidelity maxima and negative dynamical rotation peaks persists, the shallower minima of $R(t)$ reveal reduced residence time in the particle-hole symmetric configuration due to enhanced hybridization. This behavior provides a quantitative dynamical signature of the LR to NN transition and the associated strengthening of interface-mediated coupling [Fig~\ref{MP_Fidelity}(b)].

\begin{figure}[t!]
    \includegraphics[width=1.0\linewidth ]{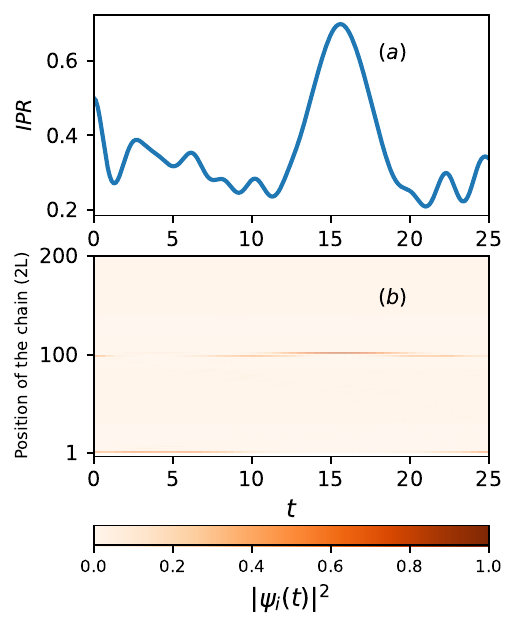}
    \caption{(a) Time evolution of the IPR. (b) Spatiotemporal profile of the time evolved state showing variation with time and site position. Parameters: $L = 100$, $\alpha = 0.7$, $J_h = 0.5$.}
    \label{heatmap_coeff}
\end{figure}

\section{Inverse Participation ratio}
\label{PR}
We consider another useful indicator of state localization, namely the \textit{inverse participation ratio} (IPR). Let the initial state of the system be
\begin{equation}
    |\psi\rangle = \sum_{i=1}^d c_i |\phi_i\rangle,
\end{equation}
where $\{|\phi_i\rangle\}$ is an arbitrary orthonormal basis and $d$ is the Hilbert space dimension.  
The IPR is defined as \cite{PhysRevB.95.115121, pg2025dependence, PhysRevE.100.042201, PhysRevB.94.155110, PhysRevA.111.052614}
\begin{equation}
    \mathrm{IPR} = \sum_{i=1}^d |c_i|^4.
\end{equation}
The IPR quantifies the concentration of the probability distribution $|c_i|^2$ in the chosen basis. For a completely localized state, where the entire weight resides on a single basis state, one obtains $\mathrm{IPR} = 1$. In contrast, for a perfectly delocalized state with uniform distribution $|c_i|^2 = 1/d$ over all $d$ basis states, the value reduces to $\mathrm{IPR} = 1/d$. Hence, a larger IPR signifies stronger localization, while a smaller IPR corresponds to greater delocalization across the basis states.

To investigate the dynamical properties of the zero-energy mode, we computed the time dependent IPR by fixing the LR exponent at $\alpha = 0.7$, a regime in which the fidelity of the initial state attains its maximum. The IPR exhibits a pronounced peak at $t \approx 16$, coinciding with the time of getting maximum fidelity and minimum dynamical rotation (compare Fig.~\ref{heatmap_coeff}(a) with Fig.~\ref{MP_Fidelity}(a)). This concurrence demonstrates that enhanced  (relatively) dynamical coherence is directly associated with the strong localization of the MZM [Fig.~\ref{heatmap_coeff}(a)].

However, the temporal evolution of the initial state alone does not resolve the spatial distribution of the MZM state. To capture this, we analyzed the spatiotemporal behavior of the initial state along the chain. At $t=0$, the MZMs are sharply localized at the edges of the left segment chain. As time progresses, the MZMs initially localized at the left segment edges gradually delocalizes and propagates across the inter segment interface. Around $t \simeq 16$, this one MZM relocalizes in right segment at left edge, signaling its  transfer into the right segment of the chain, however, the second MZM is delocalized across the chain during the dynamics. This dynamical relocation provides direct evidence of controlled, transport of zero-energy excitations across the hybrid interface [Fig.~\ref{heatmap_coeff}(b)].

The heat map confirms that, at $t \approx 16$, one MZM is clearly localized in the right segment, while the second mode remains delocalized, establishing a robust correspondence between wavefunction localization, coherence, and quantum state fidelity, which serves as a dynamical signature of Majorana-mediated quantum information transfer.

\section{Conclusion}
\label{Conclusion}

We have investigated both the static and dynamic properties of a hybrid Kitaev chain composed of a left segment with NN interaction and a right segment with LR interaction, connected via a tunable interface coupling. Our analysis of energy spectra reveals the manner in which the interplay between NN and LR interactions gives rise to MZMs and MDMs, along with their localization and energies, which are sensitively dependent on the LR exponent $\alpha$ and interface coupling strength $J_h$.

Focusing on dynamics, we monitored the fidelity, dynamical rotation, and IPR to respectively characterize the transport and localization properties of MZMs. Our results demonstrate that at $t \approx 16$, the maximum  fidelity, minimum dynamical rotation and maxima of IPR coincide, establishing a clear dynamical signature of  MZM transfer. While one MZM clearly relocates from the left to the right segment, the second mode remains dynamically dormant, likely due to its strong localization at the right edge of the target state. This observation highlights the intricate connection between localization, coherence, and fidelity, providing a symmetry-resolved framework for understanding MZM dynamics.
\par
Overall, the hybrid Kitaev chain provides a versatile platform for exploring tunable topological phenomena, enabling control over the number, energy, and spatial localization of MZMs through LR interactions and interface coupling. Specifically, the interface hopping strength and the LR exponent offer precise tunability of MZM properties. Recent progress points to feasible hybrid platforms: quantum-dot arrays permit site-resolved manipulation of proximity-induced superconducting pairing and topological phases \cite{PhysRevLett.112.086401}, while topolectrical circuits allow the effective hopping amplitudes to be engineered via adjustable circuit elements such as capacitors and resistors, providing real-time control over topological parameters \cite{PhysRevB.101.035109}. These advances indicate that hybrid, tunable Kitaev chain like systems are experimentally realistic. Taken together, our results highlight a pathway for controllable edge-state manipulation and  Majorana-mediated quantum information transfer, with promising applications in topological quantum computation.

The present work employs a sudden quench to trigger the transfer and hybridization of MZMs, offering clear insight into their intrinsic dynamical behavior. Although this non-adiabatic protocol successfully captures the essential physics of MZM dynamics. A possible extension would be to introduce a time-dependent coupling $J_h(t)$ with a tunable quench rate $\tau$, allowing the system to evolve more closely along its instantaneous eigenstates and thereby suppress unwanted excitations. Such gradual tuning could enable highly predictable hybridization and controlled transport of MZMs in realistic settings. In this context, Ref.~\cite{PhysRevResearch.6.033314} investigates adiabatic ramp schemes as a complementary technique. Their approach highlights an alternative control framework that could be combined with our quench-based method in future studies, without diminishing the relevance or generality of the dynamical insights obtained here. Taken together, the sudden-quench framework developed in this work establishes a robust platform for understanding MZM dynamics and naturally invites future extensions into more controlled adiabatic regimes.

\section*{DATA AVAILABILITY}
\label{DATA}
All the data supporting the findings in this study have already been displayed in the graphs appearing in the manuscript.

\bibliography{Kitaev}

\appendix
\section{Energy Spectrum of LR-LR Hybridization}
\label{LR_LR_hybrid}

In the main text, we focused on a hybrid Kitaev chain in which a NN segment is connected to a LR segment, which can be seen as a limiting case of a more general LR-LR hybrid chain. Here, we discuss the energy spectrum and the emergence of MDMs in this LR-LR hybridization scenario, considering two representative choices of the interface coupling.


We consider a chain of $L$ sites divided into two segments. The left segment contains $L_1$ sites, while the right segment has $L_2 (= L - L_1)$ sites. The total Hamiltonian of the hybrid LR-LR system is the sum of three contributions: the LR Hamiltonian of the left segment ($\hat H_{\rm LR,L}^h$), the LR Hamiltonian of the right segment ($\hat H_{\rm LR,R}^h$), and the interface coupling term ($\hat H_{\rm I}^h$) connecting the two segments, mathematically given as
\begin{equation}
\hat H^h = \hat H_{\rm LR,L}^h + \hat H_{\rm LR,R}^h + \hat H_{\rm I}^h.
\label{total_Ham_hybrid_LR}
\end{equation}

The left LR segment consisting of sites $1$ to $L_1$, is given by
\begin{align}
\hat H_{\rm LR,L}^h &= 
- J \sum_{j=1}^{L_1-1} \left( c_j^\dagger c_{j+1} + \mathrm{H.c.} \right)
- \mu \sum_{j=1}^{L_1} \left( c_j^\dagger c_j - \frac{1}{2} \right) \nonumber \\
&\quad + \Delta \sum_{j=1}^{L_1-1} \sum_{l=1}^{L_1-1-j} \frac{1}{d_l^{\beta}} \left( c_j c_{j+l} + \mathrm{H.c.} \right),
\label{H_LR_left_hybrid}
\end{align}
where $\beta$ defines the decay exponent of the LR pairing in the left segment. 
The right LR segment, consisting of sites $L_1+1$ to $L$, is given by
\begin{align}
\hat H_{\rm LR,R}^h &= 
- J \sum_{j=L_1+1}^{L-1} \left( c_j^\dagger c_{j+1} + \mathrm{H.c.} \right)
- \mu \sum_{j=L_1+1}^{L} \left( c_j^\dagger c_j - \frac{1}{2} \right) \nonumber \\
&\quad + \Delta \sum_{j=L_1+1}^{L-1} \sum_{l=1}^{L-1-j} \frac{1}{d_l^{\alpha}} \left( c_j c_{j+l} + \mathrm{H.c.} \right),
\label{H_LR_right_hybrid}
\end{align}
where $\alpha$ defines the decay exponent of LR pairing in the right segment. The interface Hamiltonian connecting the two segments at sites $L_1$ and $L_1+1$ is
\begin{equation}
\hat H_{\rm I}^h = J_h \left( c_{L_1}^\dagger c_{L_1+1} + c_{L_1}^\dagger c_{L_1+1}^\dagger + \mathrm{H.c.} \right),
\label{H_Interface_hybrid_app}
\end{equation}
with $J_h$ representing the interface coupling strength across the segments. 

\begin{figure}[t]
\includegraphics[width=0.49\linewidth,height=0.40\linewidth]{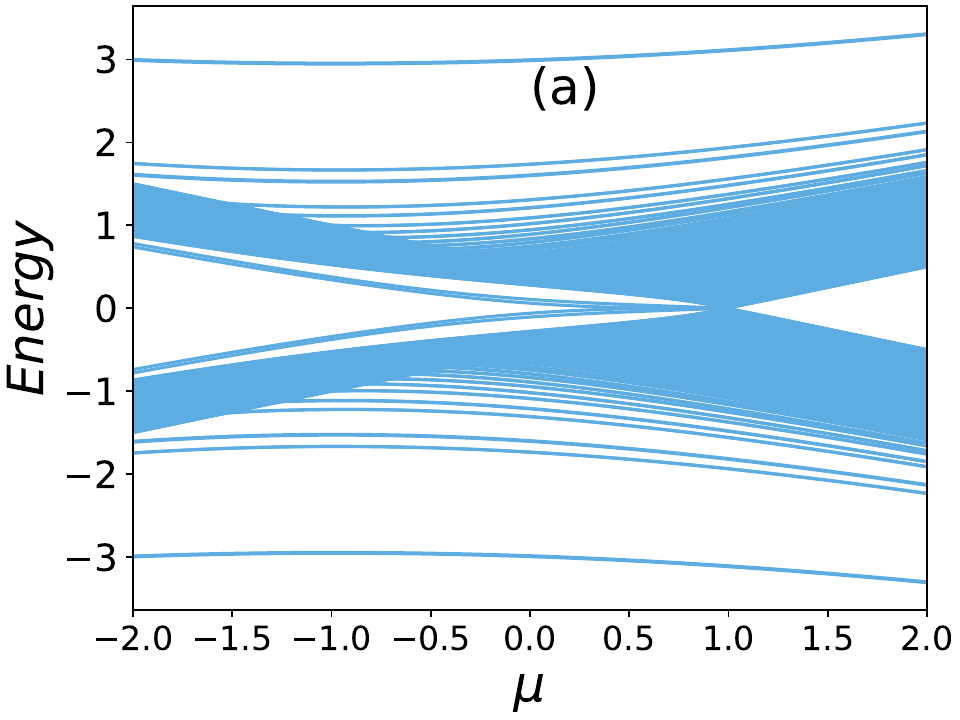}
\includegraphics[width=0.49\linewidth,height=0.40\linewidth]{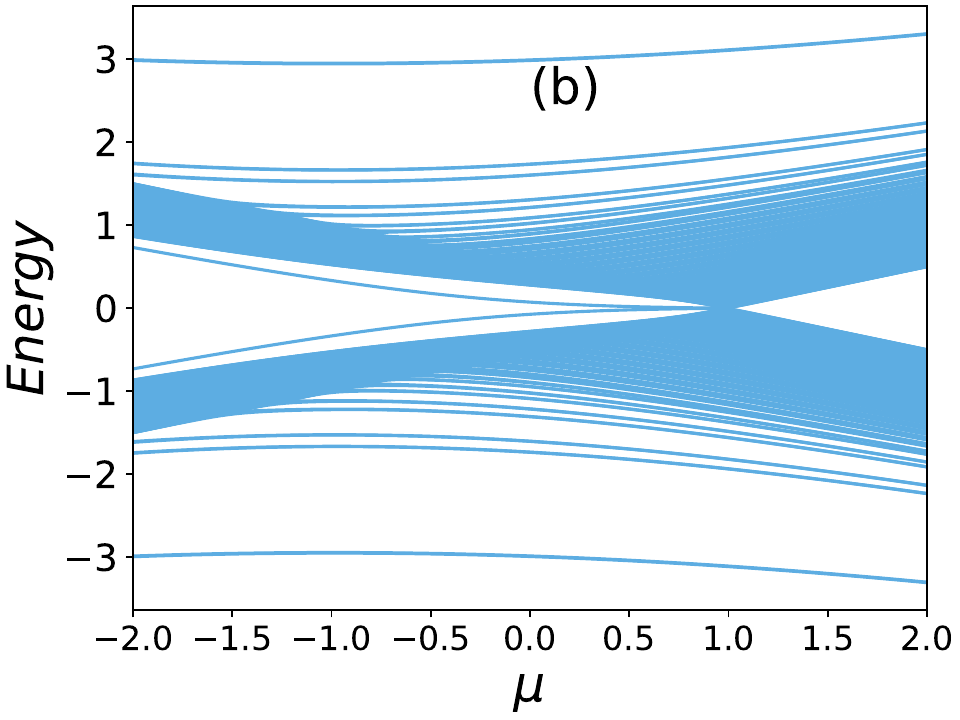}
\caption{{Energy spectrum of a LR - LR hybrid Kitaev chain composed of a LR segment ($1 \leq j \leq L_1$) and again a LR segment ($L_1 + 1 \leq j \leq L$), with $L_1 = 50$, total system size $L = 100$, under OBC. The system parameters are $J = \Delta = 1.0$. The spectrum is shown as a function of the chemical potential $\mu$ for decay exponents: $\beta = 0.5$ and $\alpha = 0.5$ and for different interface coupling strength (a) $J_h = 0.0$ and (b) $J_h = 1.0$.}}
\label{Energy_app1}
\end{figure}

We examine the energy spectrum of the LR–LR hybrid Kitaev chain for two different interface coupling strengths, fixing the LR exponents to $\beta = \alpha = 0.5$, as shown in Fig.~\ref{Energy_app1}.  When the interface coupling strength is weak ($J_h \to 0$), the chain effectively behaves as two independent LR chain, each supporting two MDMs localized at its edges. In this limit, the system hosts four MDMs in total, with their spatial decay governed by the respective exponents $\beta$ and $\alpha$ [Fig.~\ref{Energy_app1}(a)].

By contrast, when the interface coupling strength is strong ($J_h \to 1$), the two segments effectively merge into a effectively LR Kitaev chain. In this regime, only two MDMs remain, localized at the edges of the full chain, with the decay exponent unified as $\beta = \alpha = 0.5$ [Fig.~\ref{Energy_app1}(b)].

\section{Energy spectrum for asymmetric interface couplings}
\label{Interface_coupling}

\begin{figure}
\includegraphics[width=0.49\linewidth,height=0.40\linewidth]{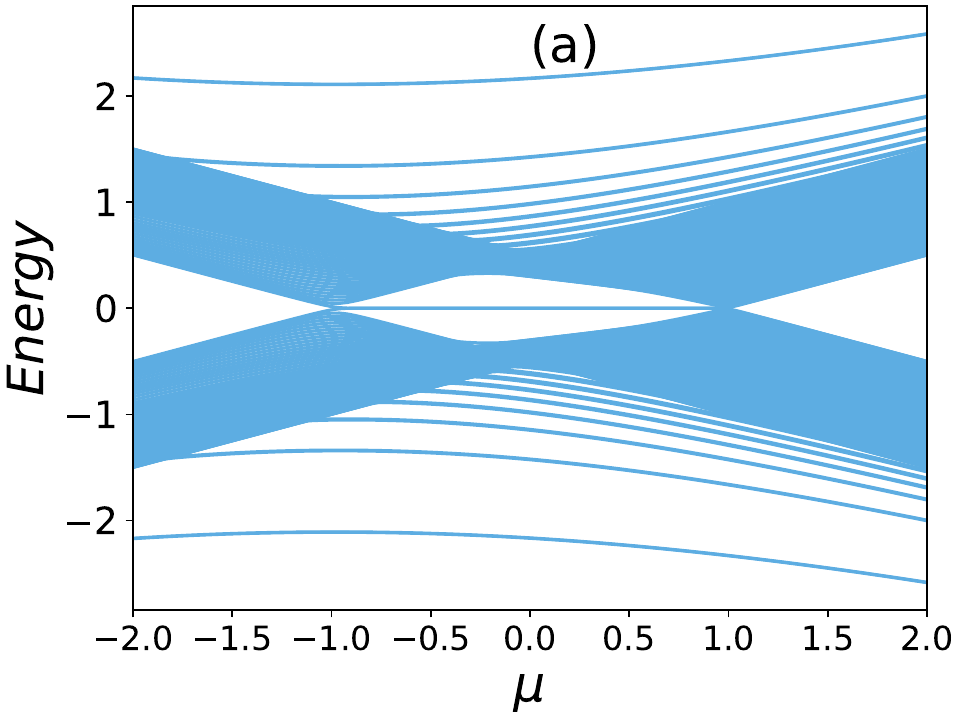}
\includegraphics[width=0.49\linewidth,height=0.40\linewidth]{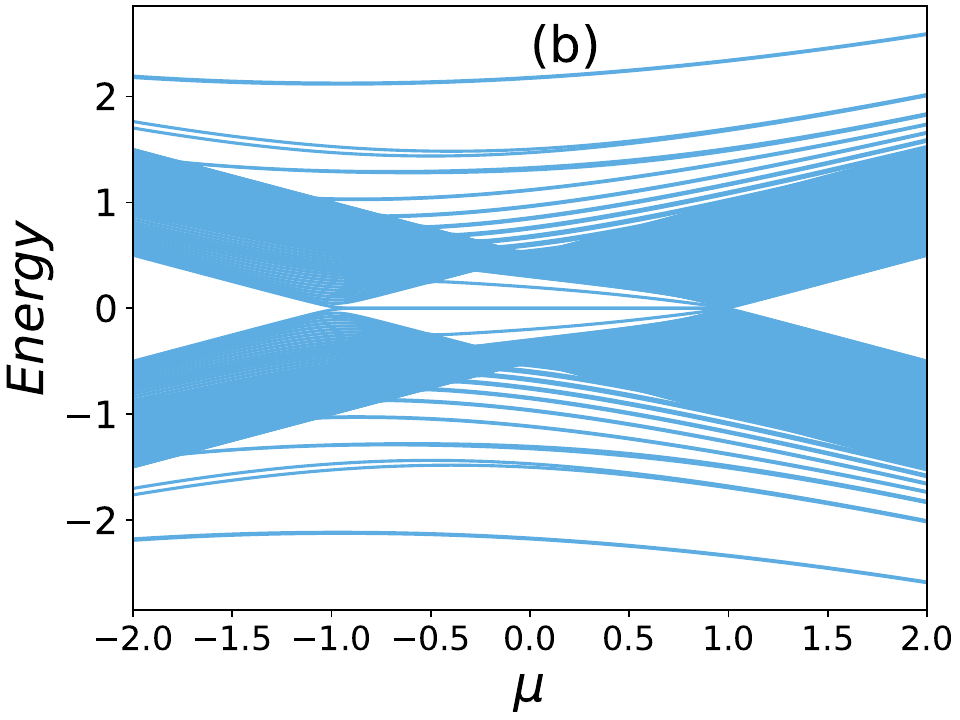}
\caption{ Energy spectrum of a NN - LR hybrid Kitaev chain composed of a NN segment ($1 \leq j \leq L_1$) and again a LR segment ($L_1 + 1 \leq j \leq L$), with $L_1 = 50$, total system size $L = 100$, under OBC. The system parameters are $J = \Delta = 1.0$. The spectrum is shown as a function of the chemical potential $\mu$ for decay exponents: $\alpha = 0.5$ for different interface coupling hoping and pairing strength (a) $J_h = 0.5$ and $\Delta_h = 1.0$ and (b) $J_h = 0.5$ and $\Delta_h = 5.0$. }
\label{Energy_app2}
\end{figure}

Throughout the main text, we restricted our discussion to the symmetric case in which the hopping and pairing amplitudes at the interface sites are identical, $J_h = \Delta_h$. While this special choice simplifies the analysis, realistic interface between a NN segment and a LR segment can naturally host different strengths for these two processes. To clarify how such asymmetry influences, and we now examine the general situation where the interface hopping and pairing are independent.

The most general bilinear interface coupling term connecting the right edge of the NN segment to the beginning of the LR segment can be written as
\begin{equation}
\hat{H}_{\mathrm{I}}^h =
J_h \left( c_{L_1}^\dagger c_{L_1+1} + \mathrm{H.c.} \right)
+
\Delta_h \left( c_{L_1}^\dagger c_{L_1+1}^\dagger + \mathrm{H.c.} \right),
\label{H_GC}
\end{equation}
where $J_h$ is hopping term across the interface, whereas $\Delta_h$ encodes the  pairing term between the segments.

Using this general interface coupling term Eq.~\eqref{H_GC} , we compute the energy spectrum for $\alpha = 0.5$ and analyze how different choices of $(J_h, \Delta_h)$ reshape the energy spectrum and analyze transition point and effect on Majorana bound states. For a mild asymmetry—specifically, $J_h = 0.5$ and $\Delta_h = 1.0$, the resulting spectrum [Fig.~\ref{Energy_app2}(a)] retains the characteristic features of the symmetric case shown earlier in Fig.~\ref{Energy J_h = 1.0}(a). The states remain well separated, and the topological region continues to host only the expected MZMs.

A qualitatively different picture emerges when the imbalance becomes large. For example, setting $J_h = 0.5$ and $\Delta_h = 5.0$ produces the spectrum in Fig.~\ref{Energy_app2}(b), where additional finite-energy states migrate into the gap even for chemical potentials $\mu < J$. The strong pairing term effectively pulls more quasi-particle levels toward low energies, giving rise to dense subgap structure coexisting with the zero-energy modes.

Crucially, this restructuring of the energy spectrum does not alter the underlying topology. The phase boundaries separating trivial and topological regions remain fixed, and the presence of MZMs is unaffected by the degree of asymmetry in $(J_h, \Delta_h)$. The interface couplings significantly modify the finite-energy landscape but leave the topological invariants and the existence of MZMs intact, fully consistent with the general form of Eq.~\eqref{H_GC}.

\end{document}